\documentclass[aps,prd,showpacs,nofootinbib]{revtex4}

\usepackage{amssymb}

\usepackage{amssymb}

\newcommand{\bls}[1]{\renewcommand{\baselinestretch}{#1}}

         \def\nqq{\hspace*{-2em}}
      
\def\cm{\hspace*{1cm}}

\def\Jl#1#2{#1 {\bf #2},\ }

\def\ApJ#1 {\Jl{Astroph. J.}{#1}}       \def\CQG#1 {\Jl{Class. Quantum Grav.}{#1}}
\def\DAN#1 {\Jl{Dokl. AN SSSR}{#1}}     \def\GC#1 {\Jl{Grav. Cosmol.}{#1}}
\def\GRG#1 {\Jl{Gen. Rel. Grav.}{#1}}   \def\JETF#1 {\Jl{Zh. Eksp. Teor. Fiz.}{#1}}
\def\JETP#1 {\Jl{Sov. Phys. JETP}{#1}}  \def\JHEP#1 {\Jl{JHEP}{#1}}
\def\JMP#1 {\Jl{J. Math. Phys.}{#1}}    \def\NPB#1 {\Jl{Nucl. Phys. B}{#1}}
\def\NP#1 {\Jl{Nucl. Phys.}{#1}}        \def\PLA#1 {\Jl{Phys. Lett. A}{#1}}
\def\PLB#1 {\Jl{Phys. Lett. B}{#1}}     \def\PRD#1 {\Jl{Phys. Rev. D}{#1}}
\def\PRL#1 {\Jl{Phys. Rev. Lett.}{#1}}

                  \def\lal{&&\nqq {}}
\def\eq{Eq.\,}                  \def\eqs{Eqs.\,}
\def\beq{\begin{equation}}      \def\eeq{\end{equation}}
\def\bear{\begin{eqnarray}}     \def\bearr{\begin{eqnarray} \lal}
\def\ear{\end{eqnarray}}        \def\earn{\nonumber \end{eqnarray}}
\def\nn{\nonumber\\ {}}         
\def\nnn{\nonumber\\ \lal }     
\def\yy{\\[5pt] {}}             \def\yyy{\\[5pt] \lal }

\def\dst{\displaystyle}             \def\tst{\textstyle}
\def\fracd#1#2{{\dst\frac{#1}{#2}}} \def\fract#1#2{{\tst\frac{#1}{#2}}}
\def\Half{{\fracd{1}{2}}}           \def\half{{\fract{1}{2}}}

                       \def\d{\partial}
\def\re{\mathop{\rm Re}\nolimits}       \def\im{\mathop{\rm Im}\nolimits}
     
\def\sign{\mathop{\rm sign}\nolimits}   
\def\dim{\mathop{\rm dim}\nolimits}     \def\const{{\rm const}}
\def\eps{\varepsilon}                   \def\ep{\epsilon}

\def\mn{_{\mu\nu}}

\def\oR{{\overline R}}      
\def\og{{\overline g}}      \def\kappa{\varkappa}

\def\CC{{\mathbb C}}        
\def\M{{\mathbb M}}         \def\R{{\mathbb R}}
\def\V{{\mathbb V}}
\def\mD{m_{{}_{\rm D}}}

\newcommand{\pd}{\d}

\bls{1.1}
\begin{document}

\title{Extra dimensions as a source of the electroweak model}

\author{S.V. Bolokhov}
\affiliation
    {PFUR, 6 Miklukho-Maklaya St., Moscow 117198, Russia.
    E-mail: bol-rgs@yandex.ru}

\author{K.A. Bronnikov}
\affiliation
    {Center for Gravitation and Fundamental Metrology, VNIIMS, 46 Ozyornaya
    St., Moscow, Russia; Institute of Gravitation and Cosmology, PFUR, 6
     Miklukho-Maklaya St., Moscow 117198, Russia.//
     E-mail: kb20@yandex.ru}

\author{S.G. Rubin}
\affiliation
    {National Research Nuclear University "MEPHI"\ , 31 Kashirskoe Sh.,
    Moscow 115409, Russia.//
    E-mail: sergeirubin@list.ru}

\begin{abstract}
    The Higgs boson of the Standard model is described by a set of
    off-diagonal components of the multidimensional metric tensor,
    as well as the gauge fields. In the low-energy
    limit, the basic properties of the Higgs boson are reproduced,
    including the shape of the potential and interactions with the gauge
    fields of the electroweak part of the Standard model.

\end{abstract}
\maketitle

\section{Introduction}

  The Standard model (SM) of particle physics is a basis of modern physics.
  Its certain shortcomings may probably be removed without essentially
  changing the basic structure of the theory. However, the origin of the SM
  is still remaining problematic. Why does the theory possess this
  particular symmetry $SU(2)\otimes U(1)$ in the present case?
  What is the origin of the Higgs field, gauge fields and matter fields?
  Why precisely three generations of fermions are realized? There are no
  unambiguous answers to these questions. The problems that exist even in
  the supersymmetric version of the SM make it necessary to extend the SM in
  search for an acceptable version \cite{Carena}.

  On the other hand, multidimensional gravity provides broad opportunities
  in explaining diverse phenomena, such as the inflaton potential and a
  nonzero value of the cosmological constant \cite{Our_Infl}. The models
  built on the basis of multidimensional gravity are intrinsically
  consistent because they contain a stabilization mechanism for extra
  dimensions \cite{Our}. Even at a phenomenological level, addition of
  nonminimally coupled Higgs fields to the Ricci scalar leads to
  nontrivial effects. As was shown in \cite{Bezrukov}, one could
  associate the inflaton and the Higgs field in this case.

  In addition, it is well known that gauge fields, including those of the
  SM, are introduced in a natural way on the basis of compact extra spaces
  (see, e.g., \cite{Blagojevic, Montani}). The off-diagonal components of
  the metric tensor are interpreted at low energies as gauge fields
  belonging to the algebra of the symmetry group of the extra space.
  The gauge symmetry of the Lagrangian thus follows from the symmetry of
  extra dimensions. Thus there are clear indications that extra dimensions
  can be a basis for the SM. In this paper, we study the possibility of
  constructing the Higgs sector of the SM, interacting with gauge fields, by
  choosing the proper geometry of the extra factor space in the spirit of
  the Kaluza-Klein approach.

  We will suggest a way of geometrization for not only the gauge fields but
  also the Higgs bosons, which transform according to the fundamental
  representation of the gauge group. As in the popular scheme of Higgs-gauge
  unification \cite{Gauge-Higgs Unif}, we identify the Higgs field with
  nondiagonal components of the metric, but in a different manner. In this
  way we are able to obtain the standard Higgs field Lagrangian containing
  interaction with the gauge fields and corresponding to the structure of
  the boson sector of the SM in the low-energy limit. The fermion sector
  is not discussed.

  Some previous papers devoted to multidimensional unified theories contain
  attempts to describe the Higgs field geometrically. In particular, in
  \cite{Vladimirov1, Vladimirov2, Vladimirov3},
  the possible introduction of an effective Higgs field was studied in the
  framework of 6-, 7- and 8-dimensional models. A characteristic feature of
  this approach was a sufficient economy of the number of extra dimensions
  in the qualitative reproduction of the gauge theories of physical
  interactions. This economy was, however, achieved at the expense of
  introducing complex quantities into the multidimensional metric and
  employing additional degrees of freedom connected with conformal (Weyl)
  transformations of the original manifold.

  In \cite{Volobuev, Volobuev2}, the authors considered the conditions for
  emergence of an irreducible scalar field multiplet due to dimensional
  reduction in multidimensional Kaluza-Klein models. The conditions found
  are of {\it sufficient\/} nature and were formulated in the framework of a
  very interesting but sophisticated mathematical scheme on the basis of
  purely group-theoretical considerations, using the method of intertwining
  operators and a generalization of Dynkin's diagram technique. Let us also
  note that the scheme used by these authors is arranged for the case of
  free gauge fields with a particular choice of the extra manifold topology
  in the form of the factor space $G/H$, where $G$ is the gauge group and
  $H$ its stationary subgroup.

  Geometrization of the Higgs sector of the SM in the framework of the
  Kaluza-Klein approach was also performed in the recent paper \cite{Niemi}.
  This model used a specific geometric structure (3-brane) in the
  background of a nontrivial geometry of extra dimensions (squashed
  three-sphere) as well as an alternative mass generation mechanism for
  vector bosons using fluctuations of the brane.

  In the present paper, we try to follow the technically simplest geometric
  approach in which the initial Lagrangian does not contain any dynamic
  variables other than the metric tensor of the multidimensional manifold.
  We do not invoke additional structures like branes or assumptions about
  some specific complicated extra space geometry. In this approach, there is
  no necessity to analyze any sophisticated group structure of the manifold
  or to introduce any non-metric degrees of freedom.

  This approach allows one not only to reproduce the conventional form of
  the Higgs sector of the SM but also to find deflections from it. The
  latter can be of particular value due to the soon expected work of the LHC
  at its full capacity and a possible negative result in search for Higgs
  particles in the predicted mass range.

\section{Statement of the problem. The metric structure of the space $\M^D$}

\subsection{Preliminaries}

  Let there be a $D$-dimensional Riemannian manifold $\M^D$ with the metric
  tensor of the form
\bear\label{totalmetric}
    (g_{AB}) = \left(
  \begin{array}{c|c|c|c}
        \og_{\mu\nu}(x,y) & g_{\mu a}(x,y) &  &   \\ \hline
        g_{b\nu}(x,y) & g_{ab}(y) &  g_{a9}(x,z) &  \\ \hline
                     & g_{9b}(x,z) & g_{99}  & \\ \hline
                        & & & g_{mn}(z),
  \end{array}\right)
\ear
  where only the most important (for this discussion) metric components
  are written. Contributions of other components to the action are not
  considered. Here and henceforth, the indices assume the following values:
\begin{description} \itemsep 0pt
\item[{}]
    $A, B, \ldots = 1, \ldots, D$ (the full dimension is $D = 9+d_3$);
\item[{}]
    $\alpha, \beta, ..., \mu, \nu,... = 1,...,4$ (the quantities
    $\og_{\mu\nu}(x,y)$ contain the metric of our 4D space
    $\M_1$ $g\mn(x) = g\mn^{(1)}(x)$, see Section IIIB);
\item[{}]
    $a, b, ... = 5,...,8$ (the notation
    $g_{ab} = g_{ab}^{(2)}= \gamma_{ab}$ will also be used;
    the space with this metric will be called $\V^4$);
\item[{}]
    $I, J, K = 5,...,9$ (the subspace with the metric $g_{IK} = g_{IK}^{(2)}$
    will be denoted $\M_2$);
\item[{}]
    $m, n, ... = 10,..., 9+d_3$ (the subspace with the metric
    $g_{mn} = g_{mn}^{(3)}$ will be denoted $\M_3$);
\item[{}]
    $i,j,k$ are used as group indices.
\end{description}
  The sets of coordinates of the subspaces $\M_1$, $\M_2$ (which includes
  $\V^4$) and $\M_3$ will be denoted by $x, y, z$, respectively, and all
  $D$ coordinates jointly by the letter $Z$.

  Let us, anticipating, describe the physical meaning of some of the metric
  components. The off-diagonal components $g_{a\mu}$, $a =5,...,8$ are
  connected, according to the standard Kaluza-Klein scheme \cite{Blagojevic},
  with gauge fields that lie in the algebra of the symmetry group of the
  4D compact factor space $\V^4$. In what follows it will be shown that
  the components $g_{a9}$ contain scalar (in $\M_1$) fields to be interpreted
  as the Higgs field. The metric tensor $g_{mn}$ of the factor space $\M_3$
  is of auxiliary nature and can in principle create inflaton-like scalar
  fields.

  The basic element in this approach is the extra space $\V^4$ with the
  metric $\gamma_{ab}$. It is supposed that its isometry group $T$
  causes the symmetry of the SM Lagrangian under gauge transformations.

  One usually does not discuss the reasons for a high symmetry of an extra
  factor space. We will also adopt this fact without proof but refer to
  the article \cite{Rubin09}, where such a reason is discussed. Briefly,
  an extra factor space is initially non-symmetric, but due to entropy
  transfer to the basic space it passes into a symmetric state. The entropy
  transfer happens due to decay of Kaluza-Klein excitations.

  Our strategy will consist in choosing the group $T$ in such a way as to
  provide the correct transformation law of the effective Higgs doublet
  that emerges from the metric components $g_{a9}$ and belonging to the
  fundamental representation\footnote
     {In the context of the present paper, a fundamental (\textit{syn.}:
     standard, or definitive) representation of a matrix Lie group is
     understood as the exact representation of minimum dimension,
     corresponding to the standard group-theoretic nomenclature specifying
     this matrix group. Thus, the fundamental representation of the
     $SU(n)$ group is realized by $n \times n$ matrices with
     determinants equal to unity (the Special Unitary group).}
  of the electroweak gauge group $SU(2)\times U(1)$ of the SM.

  The model dynamics will be specified by the $D$-dimensional action
\beq               \label{act}
    S =  \Half \mD^{D-2} \int \sqrt{^D g}\,d^D Z
        \biggl[R_D + \eta R_D^2 -2\Lambda \biggr]
\eeq
  where $R_D$ is the scalar curvature of the Riemannian space
  $\M^D$, $\eta$ and $\Lambda$ are constants (parameters of the theory),
  $\mD$ is the $D$-dimensional Planck mass. The legality of introducing
  curvature-nonlinear corrections to the action is validated
  by taking into account quantum field effects \cite{GMM,BD}. Higher-order
  corrections other than $R^2$ are not considered here because the action
  (\ref{act}) is already sufficient for the purposes of the present paper.

\subsection{Singling out the proto-Higgs field from the extra-space metric}

  Consider a class $T$ of linear transformations of the coordinates $y^a$
  of the extra space $\V^4$:
\beq\label{Ttrans}
        y'^a =T^a_{\;b} y^b, \qquad a,b=5,...,8.
\eeq
  Let us assume that $T$ forms an $s$-parametric Lie group of isometries
  of $\V^4$, i.e., these transformations leave its metric form-invariant:
\[
    g_{ab}'(y)= g_{ab}(y).
\]

  The generators $t^{ja}_{\;\;b}$ $(j=1,...,s)$ of this group,
  describing infinitesimal shifts,
\beq\label{ss}
    y'{}^a = (\delta^a_b + i\eps t^a_b) y^b,\qquad
        \eps t^a_b \equiv \sum_{j} \eps_j t^{ja}_{\;\;b},
\eeq
  create the corresponding algebra of Killing vectors on the manifold $\V^4$.

  The matrices $t^{ja}_{\;\;b}$ are assumed to be independent of $y^a$.
  In addition, it is supposed that we work in the class of coordinate
  systems $\{y^a\}$ adjusted to the linear action of the group $T$
  (\ref{Ttrans}). It is easy to verify that these transformations which
  deal with the coordinates $y^a$ only, imply the corresponding vector
  transformation law for the metric components $g_{a9}$. For infinitesimal
  shifts we accordingly have
\beq\label{trans3}
    g'_{b9}(x,z)=\frac{{\d y^a }}{{\d y'^b}}g_{a9}(x,z)
        =(\delta^a_b-i\eps t^a_b)g_{a9}(x,z);
\eeq
  The components $g_{a9}(x,z)$ (which may be called proto-Higgs fields)
  and the coordinates $y^a$ in (\ref{Ttrans}) are transformed by the same
  representation of the group $T$ (although with mutually reciprocal
  matrices belonging to this representation). This sufficiently obvious
  fact will be important in what follows.

\subsection{The choice of the symmetry group $T$}

  Let us specify the group $T$ by requiring that (a) this group should be
  realized by coordinate transformations in the extra space, i.e., it should
  be real; (b) it should be isomorphic to the electroweak group of the SM.
  In other words, the group $T$ should be built as a realification
  of $SU(2)\times U(1)$. This can be performed by taking as a basis the
  sufficiently well-known scheme of embedding the unitary groups
  $SU(n)$ into the orthogonal groups $SO(2n)$.

  Let us introduce notations for separate components of the electroweak
  group:
\beq            \label{omegas}
    \omega_1(\phi)=e^{i\phi}\in U(1);\cm
    \omega_2(\theta_j)=A(\theta_j)+iB(\theta_j)\in SU(2).
\eeq
  Here $\phi$ and $\theta_j \; (j=1,...,3)$ are parameters of the groups
  $U(1)$ and $SU(2)$, respectively; $A=\re(\omega_2)$, $B=\im(\omega_2)$ are
  real $2 \times 2$ matrices which are the real and imaginary parts of
  an $SU(2)$ element in the fundamental representation and therefore
  satisfy the conditions
\beq \label{AB}
    A^T A + B^T B=1, \qquad A^T B - B^T A=0, \qquad
     \det(A+iB) = 1.
\eeq

  We now introduce the set of $4 \times 4$ matrices $T_1,\ T_2$:
\beq \label{G12}
    T_1=\left(
  \begin{array}{cc}
    I\cos\phi & -I\sin\phi \\
    I\sin\phi &  I\cos\phi \\
  \end{array}\right),   \cm
    T_2=\left(
  \begin{array}{cc}
    A & -B \\
    B &  A \\
  \end{array}   \right),
\eeq
  where $I$ is the unit $2 \times 2$ matrix.

  The following statements are easily verified directly, taking into account
  (\ref{AB}):
\begin{enumerate} \itemsep 0pt
\item
   Each of the sets of matrices $T_1$ and $T_2$ forms a group by matrix
   product.
\item
   The group $T_1$ is orthogonal and has a unit determinant
   (i.e., lies in $SO(4)$) and is isomorphic to the group $U(1)$.
\item
   The group $T_2$ also lies in $SO(4)$ and is isomorphic to the group
   $SU(2)$ in its fundamental representation.
\item
   Matrices of the form $T_1$ and $T_2$ commute: $T_1 T_2= T_2 T_1$.
\end{enumerate}
   A consequence of all these facts is the following statement:
\textit{All possible products of elements $T_1\cdot T_2$ form a real
  4-parametric group $T$, lying in $SO(4)$ and isomorphic to the group
  $SU(2)\times U(1)$.} A constructive meaning of this statement is made
  clear in the next section. We choose as a symmetry group of the extra 4D
  space $\V^4$ the group $T=T_1 \cdot T_2$ built above.

  It is important that the initial Lagrangian is invariant under the
  coordinate transformations belonging to the group $T$ because it is
  manifestly invariant under general coordinate transformations of the
  space $\V^4$.

  It is also easy to verify that the Euclidean metric
\beq\label{flat}
        \gamma_{ab}=\delta_{ab}
\eeq
  is symmetric under transformations of the group $T$. That is why it
  can be chosen as the metric of the basic extra space $\V^4$. The latter
  is assumed to be compact, which is easily achieved, eg., by endowing
  it with toroidal topology.

\section{A transition to the dynamic variables of the SM}

  Our immediate task is to connect the Higgs complex doublet
  $h(x) \in \CC^2$ of the SM with the metric of the extra space under
  consideration. Let us address to the sector of the gauge theory
  responsible for the Higgs field and its interaction with the gauge fields
  $A_\mu(x)$. The structure of the corresponding Lagrangian is well known:
\beq
    {\cal L}[h(x),A_\mu(x)]=(D^\mu h)^{+}(D_\mu h)-V(h).
\eeq
  Here, $V(h)$ denotes the potential providing a spontaneous symmetry
  breakdown (see Section IV) while the gauge-covariant derivative
  has the form
\beq
    D_\mu h=\pd_\mu h - i g_1 {A'^a_\mu}
        \frac{\sigma^a}{2} h- \frac{i}{2} g_2 {B_\mu} h,
\eeq
  where the sets of gauge fields {$A_\mu^a$} corresponding to the
  groups $SU(2)$ and $U(1)$ are denoted by $A'^a_\mu$ and $B_\mu$,
  respectively; $\sigma_a$ are the Pauli matrices, $g_1$ and $g_2$ are
  coupling constants..

  The Higgs doublet $h$ is transformed by the fundamental representation
  of the electroweak group $SU(2)\times U(1)$:
\beq\label{group3}
        h'=\omega_1 \omega_2 h=(A+iB)e^{i\phi}h.
\eeq

  Let us establish a relation between the metric coefficients
\beq\label{ProtoHiggs}
    g_{a9}\equiv H_a
\eeq
  and the Higgs field $h$. To do so, we express the 4-component field
  $H_a$, transformed by the group $T$, in terms of two-component columns
  $X$ and $Y$:
\beq\label{invtrans}
    H\equiv\left(
          \begin{array}{c}
            X \\
            Y \\
          \end{array}
        \right),\quad   H' = T H = T_1 T_2 H,
\eeq
  With (\ref{G12}), the explicit form of the transformation (\ref{invtrans})
  in the space of the 4-columns $H$ is
\beq
    \left(\begin{array}{c} X' \\
        Y' \\
    \end{array}\right)
    = \left(  \begin{array}{cc}
        A\cos\phi-B\sin\phi & -(A\sin\phi+B\cos\phi) \\
        A\sin\phi+B\cos\phi &  A\cos\phi-B\sin\phi \\
    \end{array}
        \right) \left(
    \begin{array}{c} X \\
            Y \\
    \end{array}\right),
\eeq
  We build the combination
\beq \label{hXY}
   \tilde{h}=X + iY, \qquad X, Y \in \R^2.
\eeq
  It is easily verified that the field $\tilde{h}$ is transformed in the
  same way as the Higgs doublet (\ref{group3}). Therefore at full rights
  we identify these two doublets, $\tilde{h}\equiv h$.

  A counterpart of the transformation of complex 2-columns $h' =\omega_1
  \omega_2h$ in the space of real 4-columns is the equivalent transformation
  $H'=T H$, which means that the representations of the groups
  $T$ and $SU(2)\times U(1)$ are isomorphic.

  Let us note that the correspondence between the fields $H$ and $h$ can be
  described by using the matrix
\beq\label{P} P=\frac{1}{\sqrt{2}}\left(
    \begin{array}{cccc}
      1 & 0 & i & 0 \\
      0 & 1 & 0 & i \\
    \end{array}
  \right)
\eeq
  with the property
\beq\label{Pp}
        PP^+ =P P_R ^{-1}=1.
\eeq

  The matrix (\ref{P}) ``projects''\ one representation of the Higgs field,
  $H=(h_1, h_2, h_3, h_4)$, onto the other,
\[
    h=\left(    \begin{array}{c}
            h_1 + ih_3 \\
            h_2 + ih_4 \\
      \end{array}    \right)
\]
  so that
\beq \label {hph}
    h=\frac{1}{\sqrt{2}}PH.
\eeq

\section{Lagrangian of the Higgs field}

  So far, our main task was to extract the group structure of the Higgs
  field and the gauge fields from the extra space metric and to find the
  proper symmetry of the extra space. Let us now outline a way of obtaining
  the Higgs field Lagrangian.

  The structure of the gauge field contribution to the Lagrangian is
  studied below, therefore here we will only consider the terms that do
  not contain the metric components $g_{a\mu}(x)$ responsible for the gauge
  fields. Assuming here and further on $g_{a\mu}(x) =0$, we obtain a
  space-time with the product structure $\M^D = \M_1\otimes \M_2 \otimes\M_3$
  and the block-diagonal metric
\beq \label{H}
    ds^2  =g_{AB}dZ^A dZ^B
          = g^{(1)}_{\mu\nu}(x)dx^{\mu} dx^{\nu}
        + g^{(2)}_{IK} (x, z) dy^I dy^K
                + g^{(3)}_{mn}(z) dz^m dz^n
\eeq
  Recall that we are considering a space-time with the metric
  (\ref{totalmetric}) and the factor space dimensions $\dim \M_1 = 4$,
  $\dim \M_2 = 5$, $\dim \M_3 = d_3$. From a comparison with
  (\ref{totalmetric}) it is seen that
\bear\label{M2}
    g^{(2)}_{IK} (x, z)=\left(
      \begin{array}{c|c}
              g_{ab} &  g_{a9}(x,z)   \\ \hline
              g_{9b}(x,z) & g_{99}
      \end{array}\right)
\ear

  Let us write down the components involved in the action (\ref{act}) in
  terms of (\ref{M2}). For the scalar curvature of the whole $D$-dimensional
  space we obtain
\bearr  \label{R11}
        R_D  = \oR_1 + \oR_{2+3} + K_H ,
\nnn
        K_H = - \nabla_\alpha {\cal X}^\alpha
                + \frac 14 g^{IK, \alpha} g_{IK, \alpha}
                - \frac 14 {\cal X}^\alpha {\cal X}_\alpha,   \cm
                {\cal X}^\alpha := g^{IK} g_{IK}{}^{,\alpha}
\ear
  where $g_{IK} = g^{(2)}_{IK} (x, z)$, while a bar over $R$ means that
  the curvature is calculated in the corresponding subspace taken
  separately. In particular, $\oR_{2,3}$ is the curvature of the subspace
  $\M_2 \otimes \M_3$, equal to
\bearr  \label{R7}
    \oR_{2+3}  = \oR_2 + \oR_3 + V_H ,
\nnn
    V_H = - \nabla_n {\cal Y}^n
            + \frac 14 g^{IK, n} g_{IK, n}
        - \frac 14 {\cal Y}^n {\cal Y}_n, \cm
            {\cal Y}^n := g^{IK} g_{IK}{}^{,n};
\ear
  here $\oR_2 =0$ since the metric $g_{IK}$ does not depend on $y^I$.

  The derivatives in $x^\alpha$, forming the expression $K_H$, contribute to
  the kinetic term of the Higgs field Lagrangian. Consider the first term
  $\int d^D x \sqrt{g_D} R_D$
  in the action integral (\ref{act}). Substituting (\ref{R11}) into it and
  singling out the total derivative $\d_\alpha$, we obtain
\beq \label{div1}
       -\sqrt{g_D} \nabla_\alpha {\cal X}^\alpha
       = -\sqrt{g_1 g_2 g_3} \nabla_\alpha {\cal X}^\alpha
       = -\sqrt{g_3} \d_\alpha \Big( \sqrt{g_1\,g_2} {\cal X}^\alpha \Big)
            + \sqrt{g_1 g_3} {\cal X}^\alpha \d_\alpha \sqrt{g_2},
\eeq
  where $g_1 = |\det (g\mn)|$, $g_2 =|\det (g_{IK})|$, $g_3 =|\det(g_{mn})|$,
  and $g_3$ does not depend on $x^\alpha$.

  The expression $V_H$ in (\ref{R7}), in which the derivatives are taken
  with respect to the coordinates $z^n$, ultimately contribute to the
  potential of the Higgs field. Similarly to (\ref{div1}), let us single out
  the total derivative $\d_m$ in the action integral:
\beq
       -\sqrt{g_D} \nabla_m {\cal Y}^m                        \label{div2}
       = -\sqrt{g_1} \d_m \Big( \sqrt{g_2\,g_3} {\cal Y}^m \Big)
            + \sqrt{g_1 g_3} {\cal Y}^m \d_m \sqrt{g_2},
\eeq
  where $g_1$ does not depend on $z^m$.

  In what follows, we work in the slow-change approximation (as compared to
  the Planck scale $\mD$) suggested in \cite{Our}, according to which
  each derivative $\d_\mu\equiv \d/\d x^\mu$ is considered as an expression
  containing a small parameter $\eps$, and in the equations of motion we
  take into account only terms of orders not higher than $O(\eps^2)$.
  No small parameter is assigned to the derivative $\d/\d z^m$. As has been
  shown in \cite{Our}, this approximation is even applicable at Grand
  unification energies (under reasonable assumptions), to say nothing of
  the more modest weak interaction scale. Owing to the above-said, we
  can restrict ourselves in the kinetic term to the expression
  (see (\ref{ProtoHiggs})) $H_{a, \alpha}  H_a^{,\alpha}$, neglecting all
  further corrections, whereas in the expression for the Higgs field
  potential we preserve all quantities up to $O(H^4)$ (under the additional
  assumption $H_a H_a \ll 1$).

  In the second term of the action (\ref{act}), containing $R_D^2$, it is
  sufficient to take the expression for $R_D$ in the linear approximation
  with respect to the quantity $H_a H_a$, now without singling out total
  derivatives and neglecting expressions with $H_{a, \alpha} H_a^{,\alpha}$
  due to the above-said about the kinetic term. Therefore, we can write
\beq
    R_D^2 \approx \biggl(-\nabla_m {\cal Y}^m
            + \frac 14 g^{IK,m}g_{IK,m}\biggr)^2.
\eeq

  Let us now express the relations obtained in terms of the
  proto-Higgs field $H_a$, involved, according to (\ref{ProtoHiggs}),
  in the expression for the metric of the factor space $\M_2$ in the form
\beq\label{Met5}
    g^{(2)}_{IK}=g_{IK} = \left(
    \begin{array}{ccccc}
            -1 & 0 & 0 & 0 & H_1 (x,z) \\
            0 & -1 & 0 & 0 & H_2 (x,z) \\
            0 & 0 & -1 & 0 & H_3 (x,z) \\
            0 & 0 & 0 & -1 & H_4 (x,z) \\
        H_1 (x,z) & H_2 (x,z) & H_3 (x,z)& H_4 (x,z) & \ep
         \end{array}\right)
            = \left(   \begin{array}{cc}
            - \delta_{ab}  &  H_a \\
                            H_b  &  1
                        \end{array}\right),
\eeq
  where $\ep = \pm 1$. It is then easy to obtain
\bearr
       g_2 = |g|, \cm  g = \ep + H_a H_a,
\nnn
       (g^{IK}) = \left(   \begin{array}{cc}
               (H_a H_b)/g - \delta_{ab}  &  H_a/g \\
                    H_b/g                 &  1/g
                \end{array}\right),
\nnn
      {\cal X}_\alpha = \frac 2g \, H_a H_{a, \alpha}, \cm
       {\cal Y}_m = \frac 2g \, H_a H_{a, m}.
\ear
  Hence in \eqs (\ref{div1}) and (\ref{div2})
\beq
      \d_\alpha \sqrt{g_2} = {\cal X}_\alpha/2
        = \frac 1g \, H_a H_{a, \alpha},\cm
      \d_m \sqrt{g_2} = {\cal Y}_m/2 = \frac 1g \, H_a H_{a, m}.
\eeq
  As a result, omitting total derivatives, we obtain
\beq          \label {R_D}
       \sqrt{g_D}\ R_D = \sqrt{g_D}\ (\oR_1 + \oR_3)
        + \sqrt{g_1 g_3}\ \frac {\sign g}{2\sqrt{|g|}}
          \Big( H_{a, \alpha} H_a^{,\alpha} + H_{a, n} H_a^{,n}\Big).
\eeq
  The terms with $H_a$ form a standard expression of the form $\half(\d H)^2
  - V(H)$, where $(\d H)^2$ is obtained from expressions with the
  derivatives $\d/\d x^\alpha$, and $V(H)$ from expressions with the
  derivatives $\d/\d z^m$.

  The expression for $R_D$ gives the correct sign (plus) of the kinetic term
  of the Higgs field if $g > 0$, therefore we choose $\ep = 1$, whence
\beq
        \frac {\sqrt{g_1 g_3}} {2\sqrt{g}}
        H_{a, \alpha} H_a^{,\alpha}
         = \Half \sqrt{g_1 g_3} H_{a, \alpha} H_a^{,\alpha}
            \biggl( 1 - \Half H_b H_b + \ldots \biggr).
\eeq
  The quantity $R_D^2$ reads
\beq
    R_D^2 \approx
    \biggl(\oR_3-\nabla_m {\cal Y}^m + \frac 14 g^{IK,m}g_{IK,m}\biggr)^2
         = \biggl(\oR_3 - 2H_a \nabla_m H_a^{,m}
        -\frac 32 H_{a,m}H_a^{,m}\biggr)^2.
\eeq
  Lastly, in the third term of the action (\ref{act}) we simply expand
  $\sqrt{g_2}$ in power series with respect to $H_a H_a$,
\beq
    \sqrt{g_2} \approx 1 + \Half H_a H_a - \frac 18 (H_a H_a)^2.
\eeq

  Substituting the resulting relations to the action (\ref{act}), we have
\bear                               \label{S5}
  S_{\rm eff} &=& \Half \mD^{D-2}
    \int_{\M^D} d^4 x\ d^5 y\ d^{d_3}z \sqrt{g_1 g_3}
        \biggl[ \oR_1  + \frac 12 H_{a,\alpha}H_a ^{,\alpha}
\nnn \quad
    + \oR_{3}\biggl(1 + \Half H_a H_a - \frac 18 (H_a H_a)^2\biggr)
    + \eta \biggl(\oR_3 - 2H_a \nabla_m H_a^{,m}
              - \frac 32 H_{a,m}H_a^{,m}\biggr)^2
\nnn \quad
    + \frac 12 H_{a,n}H_a ^{,n}\biggl(1 - \Half H_a H_a \biggr)
    - 2\Lambda \biggl(1 + \Half H_a H_a - \frac 18 (H_a H_a)^2\biggr)\biggr]
\ear
  It is now easy to pass over from the proto-Higgs field $H_a$, $a=5,6,7,8$,
  to the Higgs field $h_i$, $i=1,2$. It follows from the expression
  (\ref{hph}) that $H_a H_a =h_i ^* h_i$ and so on. Therefore the expression
  (\ref{S5}) in terms of the Higgs field has the form
\bear                                   \label{HiggsAct}
  S_{\rm eff} &=& \Half \mD^{D-2}
    \int_{\M^D} d^4 x\ d^5 y\ d^{d_3}z \sqrt{g_1 g_3}
    \biggl[ \oR_1 + \frac 12 h^* _{i,\alpha}h_i ^{,\alpha} - V(h) \biggr],
\yy
      V(h) &=& \oR_{3}\biggl(1 + \Half h_i^* h_i
                - \frac 18 (h_i^* h_i)^2\biggr)
    +\eta \biggl(R_3 - h_i^* \nabla_m h_i^{,m} h_i \nabla_m h^*_i{}^{*,m}
            - \frac 32 h_{i,m}^* h_i^{,m}\biggr)^2
\nnn \quad
     + \frac 12 h_{i,n}^*h_i ^{,n}\biggl(1 - \Half h_i^* h_i \biggr) -
        2\Lambda \biggl(1 + \Half h_i^* h_i - \frac 18 (h_i^* h_i)^2\biggl).
\ear

  Up to now, the geometry of the factor space $\M_3$ remained arbitrary. Let
  us now, for simplicity, choose $\M_3$ to be one-dimensional (so that the
  full dimension is $D=10$) and compact, i.e., a ring of a certain radius
  $b_3 = b$. Thus $z \in [0, 2\pi)$ and $g_{zz} = \zeta b^2$, $\zeta =\pm
  1$. We will also assume that $\M_2$ is compact and has the form of a torus
  of certain radius $b_2$ along all five directions.

  Let us specify $h_i (x, z)$, implying that the $z$ dependence should be
  periodic with the period $2\pi$. We choose the simplest form
\beq            \label{h-chi}
        h_i(x,z) = \chi_i(x)(a \sin z + c), \cm  a, c = \const.
\eeq

  Substituting (\ref{h-chi}) into the action integral (\ref{HiggsAct}) for
  the fields $h_a(x,z)$, we integrate it over the extra dimension $z$ and
  bring it to the form
\beq                \label{A-chi}
            A_\chi = \int d^4 x \sqrt{g_1(x)}\, L_\chi, \cm
        L_\chi =  \chi^{*,\alpha} \chi_{,\alpha} - V(\chi^* \chi).
\eeq
  In doing so, we take into account that the first term in (\ref{HiggsAct})
  should take the form of the Einstein-Hilbert action
  $\half m_4^2 \int \sqrt{g_1}\oR_1$, where
  $m_4^2 = 1/(8\pi G)$ is the Planck mass squared, so that
\beq
      m_4^2 = m_{10}^8 (2\pi)^6 b_2^5 b.    \label{m4}
\eeq
  As a result, for the kinetic term in the main approximation in the sense
  described above (recall that only this approximation is considered)
  we obtain
\beq                \label{Kin}
         \frac 18 m_4^2 (a^2 + 2c^2)\chi_i^{*,\alpha} \chi_{i,\alpha}.
\eeq
  Comparing it with (\ref{A-chi}), we obtain the normalization condition
\beq                            \label{norm}
           \frac 18 m_4^2 (a^2 + 2c^2) = 1.
\eeq
  Note that the quantities $a,b,c$ have the dimensionality of length.

  For the potential $V(\chi)$, summing the contributions from all three
  terms in (\ref{act}) and taking into account (\ref{norm}), we obtain
  the following expression:
\bearr             \label{V-chi}
    V(\chi) = m_4^2 \Lambda +
        \frac {m_4^2}{4} \biggl[\Lambda(a^2 + 2c^2)
            - \frac{\zeta a^2}{2b^2}\biggr]\chi^*_i \chi_i
\nnn \cm
      + \frac {m_4^2}{64}\biggl[
        -4\Lambda(3a^4 + 12 a^2 c^2 + 4 c^4)
        +\zeta \frac {a^2}{b^2} (3a^2 + 4c^2)
        -\eta \frac {a^2}{b^4} (51 a^2 + 32 ac + 128 c^2)
            \biggr] (\chi^*_i \chi_i)^2.
\ear

  Let us compare this expression with the standard Higgs potential
\cite{Okun} \beq                            \label{V-st}
    V = \Half \lambda^2\biggl[(\chi^*_i \chi_i)^2 - \Half v^2\biggr]^2,
\eeq
  where the vacuum mean value is $v \approx 246$ GeV while the quantity
  $\lambda \leq 1$ can be expressed in terms of the Higgs boson mass
  $m_\chi = \lambda v$. The latter is expected to be in the range
  $\sim 100 \div 300$ GeV.

  Equalizing the potentials (\ref{V-chi}) and (\ref{V-st})
  term by term, we obtain:
\bearr  \cm\cm
    \Lambda = \frac {\lambda^2 v^4}{8 m_4^2},       \label{Lam}
\yyy   \cm
    \frac {a^2}{4b^2} = \lambda^2 v^2                \label{chi-2}
            \biggl(1+ \frac{v^2}{2m_4^2}\biggr),
\yyy
    - \eta v^2\biggl(                                \label{chi-4}
       204 \frac{v^2}{m_4^2} + \frac {13 c^2 + 16 ac}{b^2}\biggr) = 1.
\ear
  In (\ref{chi-2}) it has been taken into account that according to
  (\ref{Lam}), $\Lambda >0$, therefore to have a correct sign of the
  coefficient by $\chi^*_i \chi_i$ it is necessary to put $\zeta = 1$, i.e.,
  the direction $z$ is timelike. In obtaining (\ref{chi-4}) we have used
  the relations (\ref{norm}) and (\ref{Lam}); also, since
  $v^2/m_4^2 \sim 10^{-32}$, the corresponding additions to quantities
  of the order of unity are neglected; it is also remembered that
  $\lambda$ is of the order of 1.

  Furthermore, as is clear from (\ref{norm}), the quantities $a$ and $c$
  are not greater than the Planck length by order of magnitude, and from
  (\ref{chi-2}) it is evident that $a^2/b^2 \sim 10^{-32}$, so that the
  scale of the tenth dimension $b$ should be much larger than the Planck one
  (but still within the empirical constraint $b \lesssim 10^{-17}$ cm\
   corresponding to the TeV energy scale \cite{Antoniadis}).
  Moreover, returning to the dimensionless quantities $H_a$ and $h_i$,
  one can see that according to (\ref{h-chi}) they are of the order
  $\lesssim 10^{-16}$ at $|\chi_i|$ close to their vacuum value $v/\sqrt{2}$.
  So the assumption $H_a H_a = h_i^* h_i \ll 1$, used in
  (\ref{S5}) and (\ref{HiggsAct}), is well justified.

  The Higgs field mass $m_\chi = \lambda v$, according to (\ref{norm}) and
  (\ref{chi-2}), satisfies the relation
\beq
        b^2 m_\chi^2 = 2 - c^2 m_4^2/2.
\eeq
  By choosing the parameter $c$ it can be made a quantity of the order
  of the vacuum value $v$, much smaller than  $1/b$, which
  has the order of (at least) a few TeV. Precisely this is required
  in the SM \cite{Okun}.

  Lastly, from (\ref{chi-4}) it follows that the parameter $\eta$ in this
  model should be rather a large negative quantity,
  $-\eta \sim 10^{30} v^{-2} \sim 10^{-2}\ {\rm cm}^2$. However,
  being applied to the gravitational field in our space $\M_1$, the
  correction $\eta R^2$ cannot be significant at curvatures
  $R < 1\ {\rm cm}^{-2}$, i.e., at curvature radii larger than 1 cm.

  We can conclude that the parameters of the Higgs field potential
  (\ref{V-chi}), obtained here due to a specific choice of the
  extra-dimensional metric components, agree with those involved in the SM.

\section{Interaction between the Higgs field and the gauge fields of the SM}

  We have singled out the components of the metric tensor interpreted
  as the Higgs bosons. Meanwhile, the way of singling out gauge fields
  from the extra-dimensional metric is well known. Namely, the metric
  (\ref{totalmetric}) is represented in a  standard form (see, e.g.,
  \cite{Wesson}), where the following components of the total metric
  (\ref{totalmetric}) will be of interest for us:
\bearr  \label{one}
     \og_{\mu\nu}(x,y) = g_{\mu\nu}(x) + g_{ab}k^{a}_{i}(y)A^i_{\mu}(x)
        k^{b}_{j}(y) A^j _{\nu}(x) ,\qquad i,j = 1,2,3,4
\yyy    \label{two}
     g_{\mu a}(x,y)= g_{ab}k^{a}_{i}(y)A^i_{\mu} .
\ear
  Here, $k^a_i$ is a Killing vector of the subspace $\V^4$ with the metric
  $g_{ab}$,  $A^i_{\mu}(x)$ are gauge fields in the algebra of the
  symmetry group $T$ of $\V^4$.

  The gauge field Lagrangian is obtained in the conventional way from
  a decomposition of the multidimensional scalar curvature, starting from
  the ansatz (\ref{one}) characteristic of the Kaluza-Klein theories.
  As a result, from the original multidimensional action (\ref{act}),
  after integration over the compact submanifolds, one singles out the
  effective action of the form
\beq
    S_{\rm eff}=\const \int_{\M_3} d^{d_3}z
    \int_{\M_1}d^4x\sqrt{|g_{\mu\nu}|}\left(\frac{R_4}{2\kappa}
    -\frac{1}{4}{\rm Tr}({\bf F_{\mu\nu}}{\bf F^{\mu\nu}}) \right),
\eeq
  which includes the 4D scalar curvature $R_4$ and the Lagrangian of free
  Yang-Mills fields \cite{Blagojevic}. We will no more discuss this
  well-known part of the Lagrangian. Our purpose will be a consideration
  of its part containing the Higgs field (see the previous section) and its
  interaction with the gauge fields.

  We will need the structure constants $f_{jkl}$ involved in the equation
  for the Killing vectors $k_j^a (y)$ since the same structure constants
  determine the algebra of the four gauge fields $A^i _{\mu}(x)$. To find
  the form of the structure constants, let us note that one of the possible
  representations of the Killing vectors is determined by the matrices
  $t^{j}_{ab}$ from (\ref{ss}). Indeed, a displacement along the vectors
\[
        k_j ^a (y) \equiv it^{ja}_{b}y^b
\]
  does not change the metric because $t^{ja}_{b}$ are generators of the
  symmetry group. Evidently, the nonzero structure constants $f_{jkl}$
  are the structure constants of the group $T_2$, coinciding with the
  structure constants $\eps_{jkl}$ of the $SU(2)$ group due to their
  isomorphism (see \eq (\ref{eijk}) below).

  To construct a gauge-invariant derivative $D_\mu H$ corresponding to
  the group $T = T_1\cdot T_2$ we need its generators. The generator
  $\hat t_0$ of the group $T_1$ is found directly:
\beq
    \hat t_0=\frac{\d}{\d\phi}T_1\biggr|_{\phi=0}=
    \left(
    \begin{array}{cc}
    O & -I \\ I & O
    \end{array}\right).
\eeq

  The generators $\hat t_j \, (j=1,2,3)$ of the group $T_2$ are easily
  constructed from the generators $\tau_j=-i\sigma_j/2$ of the group $SU(2)$
  satisfying the commutation relations $[\tau_j, \tau_k]=\ep_{jkl}\tau_l$,
  where $\sigma_j$ are the Pauli matrices. Since
  $\tau_j  \equiv \d \omega_2/\d\theta_j \bigr|_{\theta=0}$,
  taking into account (\ref{omegas}) and (\ref{G12}), we obtain
\beq
    \hat t_j=\frac{\pd}{\pd\theta_j}T_2\Bigr|_{\theta=0}=
    \left(\begin{array}{cc}
    \re(\tau_j) & - \im(\tau_j) \\
    \im(\tau_j) &   \re(\tau_j)
        \end{array}\right).
\eeq
  Since $\tau_2$ is real while $\tau_1$ and $\tau_3$
  are pure imaginary, we can write
\beq    \label{generators}
    \hat t_1=\left(\begin{array}{cc}
    O & i\tau_1 \\
    -i\tau_1 & O
        \end{array} \right),\qquad
    \hat t_2=\left(\begin{array}{cc}
    \tau_2 & O \\
    O & \tau_2
        \end{array} \right),\qquad
    \hat t_3=\left(\begin{array}{cc}
    O & i\tau_3 \\
    -i\tau_3 & O
    \end{array} \right).
\eeq

  It is easy to verify that the corresponding commutators coincide
  with the commutators of the $SU(2)$ group:
\beq \label{eijk}
    [\hat t_j, \hat t_k]=\ep_{jkl}\hat t_l,
\eeq
  which corresponds to isomorphism between representations of the
  groups $T_2$ and $SU(2)$.

  The structure of the term $L_{\rm int}(A,H)$, containing interaction
  between the field $H$ and the gauge fields $A_\mu$, can be obtained from
  general considerations. Since the initial Lagrangian is invariant under
  general coordinate transformations and therefore under those belonging
  to the group $T$, the fields $A$ and $H$ must enter into the Lagrangian
  in a certain gauge-invariant combination. The latter is well-known and has
  the form
\bear
      L_{\rm int}(A,H) &=& {\cal N} g^{\mu\nu}(D_\mu H)^+ (D_\nu H),
\nn
    (D_\mu H)_a &=& \left(\delta_{ab}\d_{\mu}
        +A^{i}_{\mu}(x)\hat t_{i,a b}\right)H_b,
\ear
  where ${\cal N}$ is a certain constant factor determined from the
  expression for the kinetic term of the Higgs field. The coupling constants
  and the charge factors are normalized to unity. It is convenient to
  split the set of gauge fields according to the factor groups
  $SU(2)$ and $U(1)$:
\[
    A^{j}_{\mu}(x)\hat t_{j,ab} \equiv \sum_{m=1,2,3}
    A^{m}_{\mu}(x)\hat t_{m,ab}+ B_{\mu}(x)\hat t_{0,ab}.
\]

  To pass over to the full set of dynamic variables of the SM, it is
  necessary to express the Lagrangian  $L_{\rm int}(A,H)$ in terms of
  the complex scalar doublet $h$. An explicit transition from the
  $H$-representation to the $h$-representation is given by the relations
\beq    \label{hh}
        X = \frac{1}{2}(h^{*}+h),\qquad
        Y = \frac{i}{2}(h^{*}-h),
\eeq
  easily obtainable from (\ref{hXY}) along with \eqs (\ref{generators})
  for the generators. The term $L_{\rm int}(A,H)$ then becomes a certain
  function $\tilde L_{\rm int}(A,h)$ of the fields $h$ and their derivatives:
\beq \label{HAh}
    L_{\rm int}(A^j_\mu(x), H_a(x)) = \tilde L_{\rm int}(A^j_\mu(x), h_i(x)).
\eeq

  Let us note that the above-mentioned invariance of the Lagrangian
  $L(A,H)$ under transformations from the group $T$ implies $SU(2)\times
  U(1)$ invariance of the Lagrangian $\tilde L(A,h)$. Indeed, suppose that,
  in the space of fields $H$, a $T$-transformation $H'=TH$ has been
  performed; in the space of fields $h$ it corresponds to an $SU(2)\times
  U(1)$ transformation $h'= \omega_1\omega_2h$. The fields of the pair $(H',
  h')$ are connected with each other by the same transformation $P$ from
  (\ref{hph}) as the fields of the pair $(H, h)$. Therefore $L(A',H') =
  \tilde L(A',h')$ similarly to (\ref{HAh}). On the other hand, due to the
  $T$-invariance, $L(A,H)=L(A',H')$.  Unifying these equalities into a
  single chain, we obtain $\tilde L(A',h')=L(A', H')=L(A, H)=\tilde L(A,
  h)$, whence it follows that
\beq                           \label{inv}
        \tilde L(A, h) = {\rm inv}\ [SU(2)\times U(1)].
\eeq

  The form of the expression bilinear with respect to the Higgs fields
  and invariant with respect to the group $SU(2)\times U(1)$,
  is well known:
\beq \label{Linth}
        \tilde L_{\rm int}(A,h)= {\cal N}g^{\mu\nu}(D_\mu h)^{+}(D_\nu h),
\eeq
  where ${\cal N}$ is a constant factor and
  the gauge-invariant derivative of the field $h$ has the form
\[
    D_{\mu}h \equiv (\d_{\mu} + A_{\mu}^m \tau_m + B_{\mu}I)h.
\]

  Further, by analogy with the previous part, we substitute the ansatz
  (\ref{h-chi}) into \eq (\ref{Linth}) and, integrating over
  all extra dimensions, we finally obtain
\beq \label{Linth2}
    \tilde L_{\rm int}(A,\chi)={\cal N}'
            g^{\mu\nu}(D_\mu \chi)^{+}(D_\nu \chi).
\eeq
  The normalization factor ${\cal N}' = 1$, as follows from comparison with
  the factor at the kinetic term in (\ref{A-chi}).

  The expression (\ref{Linth2}) contains the standard form of interaction
  between the gauge fields and the Higgs field belonging to the fundamental
  representation of the gauge group and corresponding to the boson sector
  structure of the SM.

\section{Discussion}

  We have shown in this paper that the boson sector of the unified weak
  and electromagnetic interactions can be reproduced in the framework of
  multidimensional gravity. In doing so, it has been sufficient to
  introduce a 5D compact extra space possessing a certain symmetry.
  It has turn out to be unnecessary to introduce such extended structures
  as strings or branes. No additional fields apart from the metric tensor
  have been used. Thus we have been able to indicate a possible purely
  geometric origin of the Higgs field.

  One of the attractive features of the idea of extra dimensions in the
  Kaluza-Klein spirit is a natural way of introducing gauge symmetries:
  the off-diagonal components of the metric play the part of gauge fields.
  We have shown that the Higgs field can also be introduced in the framework
  of this paradigm: namely, we have found such components of the
  extra-dimensional metric that can be interpreted as the Higgs field
  with a correct transformation law under the action of the group
  $SU(2)\otimes U(1)$ and a standard coupling to the gauge fields.
  The potential of the Higgs field also has its standard form.

  There are a number of ways to extend the SM in order to get rid of its
  shortcomings (see, e.g., \cite{Foot}). Our results also coincide with
  the SM in the first approximation only. Thus, the Higgs field potential
  is quite a complicated function, coinciding with the well-known expression
  only at low energies. Apart from the Higgs bosons, there are a number
  of other scalar and vector fields, which is inherent to an approach
  employing extra dimensions. Distinctions from the SM, existing in this
  model, can be used for a correct interpretation of new results
  expected at the LHC.

\subsection*{Acknowledgments}

  This work was supported in part by the Russian Foundation for
  Basic Research (project no. 09-02-00677-a) and by the Federal Program
  ``Research and Pedagogical Specialists of Innovative Russia for
  2009-2013''. SB and KB were also partly supported by NPK MU grant
  at PFUR.

\end{document}